\documentclass[%
 reprint,
superscriptaddress,
nofootinbib,
 amsmath,amssymb,
 aps,
pra,
]{revtex4-2}

\pdfoutput=1
\usepackage[utf8]{inputenc}
\usepackage[english]{babel}
\usepackage[T1]{fontenc}
\usepackage{amsmath}
\usepackage{amssymb}
\usepackage{hyperref}

\usepackage{tikz}
\usepackage{lipsum}
\usepackage{physics}
\usepackage{todonotes}

\begin{document}

\title{Halving the width of Toffoli based constant modular addition to n+3 qubits}

\author{Oumarou Oumarou}
\affiliation{Clausthal University of Technology, 38678 Clausthal-Zellerfeld, Germany}

\author{Alexandru Paler}
\affiliation{University of Texas at Dallas, Richardson, TX 75080, USA}
\affiliation{Transilvania University, 500036 Brașov, Romania}
\affiliation{Johannes Kepler University, 4040 Linz, Austria}

\author{Robert Basmadjian}
\affiliation{Clausthal University of Technology, 38678 Clausthal-Zellerfeld, Germany}

\begin{abstract}
We present an arithmetic circuit performing constant modular addition having $\mathcal{O}(n)$ depth of Toffoli gates and using a total of $n+3$ qubits. This is an improvement by a factor of two compared to the width of the state-of-the-art Toffoli-based constant modular adder. The advantage of our adder, compared to the ones operating in the Fourier-basis, is that it does not require small angle rotations and their Clifford+T decomposition. Our circuit uses a recursive adder combined with the modular addition scheme proposed by Vedral et. al. The circuit is implemented and verified exhaustively with QUANTIFY,  an open-sourced framework. We also report on the Clifford+T cost of the circuit.
\end{abstract}

\maketitle

\section{Introduction}

Arithmetic quantum circuits, namely adders, modular adders and multipliers, are an integral part of the implementation of practical quantum algorithms. In practice, adders, e.g. \cite{cuccaro2004new, draper2000addition, draper2006logarithmic, munoz2018quantum}, are building blocks of more complex functions, such as the modular exponentiation needed for Shor's algorithm \cite{beauregard2003circuit, fowler2004implementation, haner2017factoring}. Optimising adders as a sub-circuit can eventually benefit the entire circuit due to the convoluted relation that exists between the arithmetic operations. There are also quantum arithmetic circuits developed specifically for Shor's algorithm, such as \cite{beauregard2003circuit, haner2017factoring}. These exploit the fact that one of the inputs is an integer which does not require quantum storage.

Modular addition is a fundamental operation in Shor's algorithm. The circuit for modular addition takes three integer inputs $a$, $b$ and the ring size $N$, and outputs $a+b$ or $a+b-N$ depending on whether $a+b<N$ or not. In \cite{haner2017factoring}, the presented modular adder is composed of two comparators and an adder. The compartors in the modular adder circuit use approximately $2n$ qubits (i.e width) but the adder only uses around $n$ qubits. Hence, the width of the entire circuit ($\sim 2n$) is dictated by the comparators.

In \cite{vedral1996quantum}, no Comparators per se are used, but still the width of the modular adder is even greater when compared to the aforedescribed circuit because the adder used only, disregarding the other employed registers, needs $3n+1$ qubits.

Modular adders of low depth and using $\mathcal{O}(n)$ qubits are known, but use the QFT \cite{draper2000addition} approach. The QFT uses controlled rotation gates, and the angles are of the form $e^{\frac{i2\pi k}{N}}$, where the maximum value of $N$ is $2^n$ for addition on $n$ qubits. The rotation angles get smaller with increasing number of qubits. When error-correcting such circuits, the controlled rotation gates have to be decomposed into Clifford+T. The decomposition procedure introduces on the order of hundred T gates per rotation gate \cite{njross} (exact number depends on the decomposition approximation precision), such that QFT modular addition is not necessarily resource efficient when error-corrected.

The main contribution of this work is a method that performs constant modular addition using the adder from \cite{haner2017factoring} while bypassing the need for comparators. Inspired by the modular addition method from \cite{vedral1996quantum}, we combine it with the adder from  \cite{haner2017factoring} to yield a modular adder with a linear depth of $\mathcal{O}(n)$ and a qubit width of $n+3$.  Consequently, the width of constant modular addition is halved and reduced very close to its minimum of $n$ wires.

The rest of this paper is organised as: In Section \ref{sec:pre}, we present the circuits used to construct our modular adder. In Section \ref{sec:methods}, we present the design method and steps for our circuit. Finally, in Section \ref{sec:analysis} we investigate different decompositions into Clifford+T scenarios, analyse and compare them.

\section{Preliminaries}
\label{sec:pre}

In the following, we review the construction of the recursive adder from \cite{haner2017factoring}, and the modular addition method from \cite{vedral1996quantum}. 

\subsection{The Incrementer and the Carry Gates}
\label{sec:inc}

The Incrementer is a circuit that adds one to the value of an integer stored in a quantum register. Incrementation can be achieved with the help of a quantum adder with $\ket{a}$ as the operand to be incremented, while the value of the second operand $\ket{g}$ is irrelevant (garbage register). The second register is used in order to perform a trick based on the two's complement representation. We denote by $\Tilde{g}$ the bitwise negation of the number $g$. For numbers represented as two's complement, $\Tilde{g}= -g - 1$, such that $g + \Tilde{g} = -1$. Thus, the value $a$ can be incremented by performing $a-g-\Tilde{g}$. In terms of a quantum circuit, the incrementation procedure is the following:
\begin{itemize}
    \item Subtract $g$ from $a$: $\ket{a}\ket{g} \xrightarrow{} \ket{a-g}\ket{g}$
    \item Flip all the qubits of the garbage register $g$: $\ket{a-g}\ket{g} \xrightarrow{} \ket{a-g}\ket{\Tilde{g}} $ 
    \item Subtract $\Tilde{g}$ from $a-g$: $\ket{a-g}\ket{\Tilde{g}} \xrightarrow{} \ket{a-g-\Tilde{g}}\ket{\Tilde{g}} =  \ket{a-g+g + 1}\ket{\Tilde{g}}$
    \item Restate the garbage register to its original state: 
     $\ket{a+1}\ket{\Tilde{g}} \xrightarrow{} \ket{a+1}\ket{g}$
\end{itemize}

\begin{figure}[!t]
    \centering
    \includegraphics[scale=0.4]{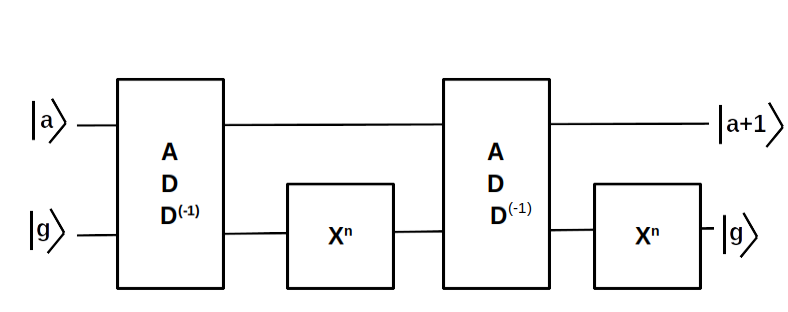}
    \caption{The Incrementer. There are two quantum registers: $a$ stores the operand, $\ket{g}$ is a garbage register. $Add^{-1}$ is the subtraction operation implemented using the inverse of an addition circuit. $X^n$ is $n$ X gates applied on the garbage register.}
    \label{fig:my_label}
\end{figure}

The Carry gate proposed in \cite{haner2017factoring} determines the most significant bit of a sum of two integers. The circuit uses a classical $n$-bit constant $c$, an $n$-qubit register $\ket{a}$ which stores the first operand, a garbage register but of $n-1$ qubits, and an ancilla initialised to $\ket{0}$. The content of the garbage register is irrelevant, but will be used during the computation. The $n^{th}$ bit of the sum $(a+c)$ will be stored in the third register as:
\begin{align*}
    \ket{a}\ket{g}\ket{0} \xrightarrow{Carry Gate} \ket{a}\ket{g}\ket{(a+c)_n}
\end{align*}

\subsection{The Recursive Adder}
\label{sub:recadd}

The recursive adder from \cite{haner2017factoring} computes the sum of two integers: the quantum register $\ket{a}$ and the classical constant $c$. This adder uses mainly two sub-circuits: the Carry gate and the Incrementer of Section \ref{sec:inc}. The inputs to the adder are a quantum register of size $n+1$, a garbage qubit $\ket{g}$ and a classical constant $c$. It outputs the sum $\ket{a+c}\ket{g}$.

The recursive adder is an in-place adder: the construction uses the fact that the sum bit at position $m$ depends on the carry bit generated by the $m-1$ bits before. For $m$ being the middle of the $\ket{a}$ bitstring, $m=\frac{n}{2}$, and knowing the carry bit from the first half of the bits (use Carry gate), the second half of the bits can be treated as a separate number which is just incremented (use Incrementer gate). For simplicity of demonstration, we consider the garbage qubit to be initialised to $\ket{0}$. Later we present the general concept with an arbitrary value of $\ket{g}$. The addition procedure is performed as:
\begin{enumerate}
    \item Split the register $\ket{a}=\ket{a_H}\ket{a_L}$ where $a_H$ and $a_L$ are respectively the higher and lower halves of the binary representation of $a$. Split also the constant $c$ in the same manner as $\ket{a}$.
    \item Apply the Carry gate to the $\ket{a_L}$ and $c_L$ using $\ket{a_H}$ as garbage register and the garbage qubit $\ket{g}$ to store the carry bit of $a_L+c_L$.
    \item Use the garbage qubit $\ket{g}$ to control whether the upper half $a_H$ should be incremented or not. If $\ket{g}=\ket{1}$ then we have a carry bit from $a_L+c_L$ and it should be added to $a_H$. Hence the Incrementer should be applied. Otherwise, we don't apply it.
    \item To reset the carry qubit to $\ket{0}$, reapply the Carry gate.
    \item Recursively apply the previous three steps to $a_L$ and $a_H$.
\end{enumerate}

\begin{figure}[!t]  
  \centering
    \includegraphics[width=\columnwidth]{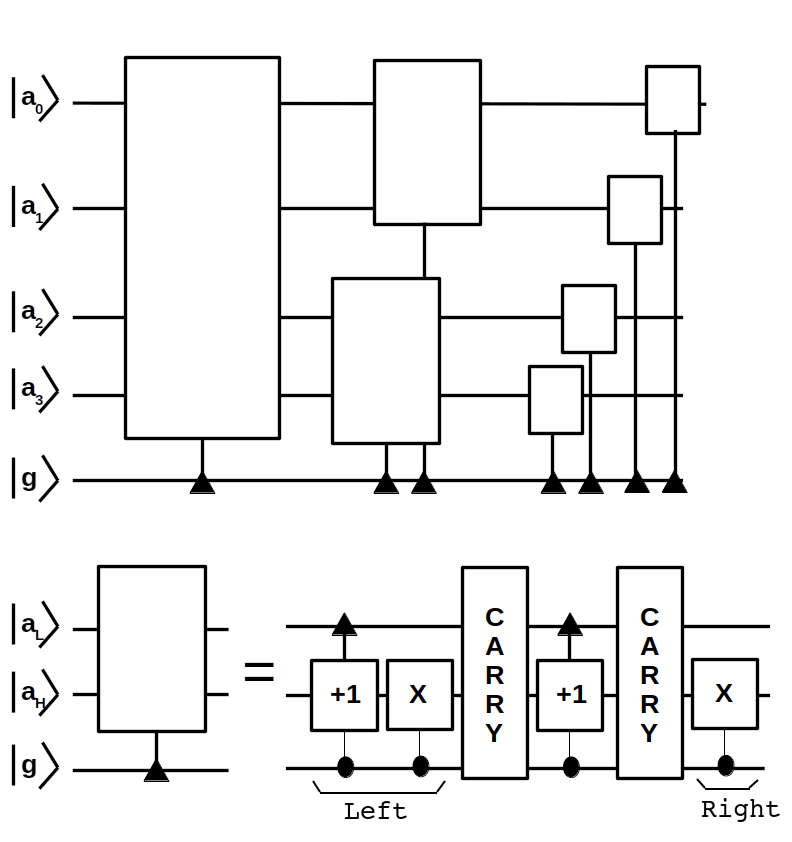}
    \caption{Recursive adder circuit design \cite{haner2017factoring}. The registers $\ket{a_L}$, $\ket{a_H}$ are the lower, upper halves of the input integer $a$ respectively and $\ket{g}$ is the garbage qubit used as a control for the Incrementer and the X gates and as a result qubit for the Carry gates. The triangles point to the used garbage qubits.}
    \label{fig:recadd}
\end{figure}

The upper part of Figure \ref{fig:recadd} illustrates the recursion process for the case of 4-bitstring  $\ket{a}$. The left most part (e.g the big box) represents the entire one 4-bitstring which is divided into two 2-bitstrings (e.g. the middle two boxes). Those are on their turn divided into four 1-bitstring (e.g. the right 4 small boxes). Note that the recursion stops when one n-bitstring is subdivided into $n$ 1-bitstrings.

The addition can work with arbitrary values stored in the garbage register by appending to the Carry-Incrementer-Carry sequence in  Figure~\ref{fig:recadd}:
\begin{itemize}
    \item Left: an Incrementer on $\ket{a_H}$ controlled by $\ket{g}$ a set of $CNOT$ gates also targeting $\ket{a_H}$ and controlled by $\ket{g}$ from the left.
    \item Right: another set of $CNOT$ gates.
\end{itemize}

This construction works because if the initial garbage is $\ket{0}$, the circuit is just as listed before. For $\ket{g}=\ket{1}$, the first Incrementer generates $\ket{g+1}$, the bitwise negation results in $\ket{\Tilde{g}}=-(g+1)-1=-g-2$. There are two options: a) the first carry flips $\ket{g}$ such that the lowest bit is $\ket{0}$, the Incrementer and the second Carry are not called, and the second negation returns the state to $\ket{g+1}$; b)  the first carry does not flip $\ket{g}$, the controlled-Incrementer is called such that $\ket{-g-1}$, the second Carry is not called and the final bit flips result in $\ket{\widetilde{-g-1}} = g + 1 - 1 = g$.

\subsection{Modular Addition}

The intuitive way of constructing a modular adder is to use a Comparator gate to test the sum of the two operands with the maximum value representative in the ring. Based on the comparison result, we either only add the two operands, or, in case there was an overflow, subtract $N$ from the sum. This approach to modular addition, used by \cite{haner2017factoring}, requires $2n+\mathcal{O}(1)$ wires because of the comparisons. 

\begin{figure}[!t]
    \centering
    \includegraphics[scale=0.42]{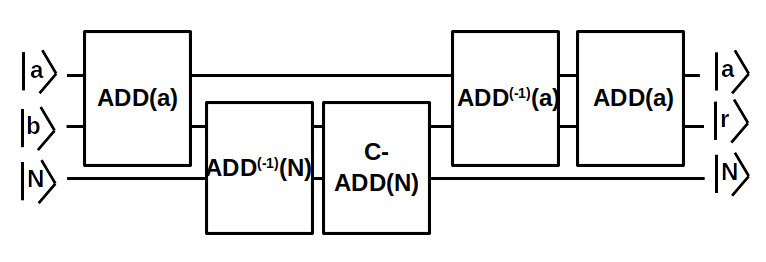}
    \caption{Modular adder circuit design used in \cite{vedral1996quantum}. $a$, $b$ are the two integers and $N$ is the size of the ring. $\ket{a}$, $\ket{b}$ are quantum registers of size $n$ and $n+1$ respectively. The addition and subtraction are performed using the adder in \cite{vedral1996quantum} requiring hence another quantum register of size $n$ for carry bits which is not depicted here.}
    \label{fig:barenco}
\end{figure}

Compared to the intuitive approach from \cite{haner2017factoring}, the modular adder in \cite{vedral1996quantum} has higher depth and an $4n+\mathcal{O}(1)$ width. However, the modular addition approach is general and not tied to a particular adder design -- the adder can be replaced. In Section~\ref{sec:methods}, we use in-place recursive adders for the additions from Figure~\ref{fig:barenco}. Also, two of the inputs, accounting for $2n$ qubits can be replaced by classical values in the case of constant addition.

We successfully designed a modular adder that uses only n+3 qubits which is approximately a $50\%$ reduction compared to state of the art modular adder like in \cite{haner2017factoring} while simultaneously maintaining the linearity in the depth.  
\section{Methods}
\label{sec:methods}
In \cite{haner2017factoring}, the Comparator circuits are the culprit behind the $2n$ width of modular addition, although the modular adder uses for the addition/subtraction only approximately half of the wires. At the same time, the modular adder from \cite{vedral1996quantum} does not use Comparator circuits, and is agnostic of how the adders are implemented. We use a recursive adder of width $n+2$ from \cite{haner2017factoring} (in Figure~\ref{fig:recadd} $\ket{a}$ is of size $n+1$ and there is an additional ancilla $g$ of size 1 qubit) to implement the modular addition. The advantage is that we halve the width because we eliminate the need for Comparator circuits.

The Comparator in \cite{haner2017factoring} is implemented by applying their Carry circuit, which uses $2n+\mathcal{O}(1)$ qubits, out of which $n$ are garbage. When implementing recursive addition, the doubling of qubits is not an issue, because in the adder half of the $n$ qubits are garbage for the other half (cf. Figure~\ref{fig:recadd}). However, for the Comparator circuit this approach is not efficient, because the Carry has to consider all $n$ qubits. Although one of the modular adder diagrams in \cite{oonishi2020efficient} shows $n$ wires, internally the adder uses $n$ ancillae for a total width of $2n$.

It is possible to implement constant-modular addition without a Comparator circuit and the corresponding ancillae \cite{vedral1996quantum}. We implement the recursive adder using: a) the Carry gate from \cite{haner2017factoring}, and b) the Incrementer is the linear-depth controlled-adder (CtrlAdd) from \cite{munoz2018quantum}. The original CtrlAdd circuit has a width of $2n+3$, and we can cut two of the ancillae because we made sure from the size of the input register $\ket{a}$ that the incrementation never overflows. The CtrlAdd circuit will be applied to only half of the bits from the recursive adder, and the other half are used as garbage \cite{munoz2018quantum} (see Figure~\ref{fig:recadd}). 

\begin{figure}[!t]
    \centering
    \includegraphics[width=\columnwidth]{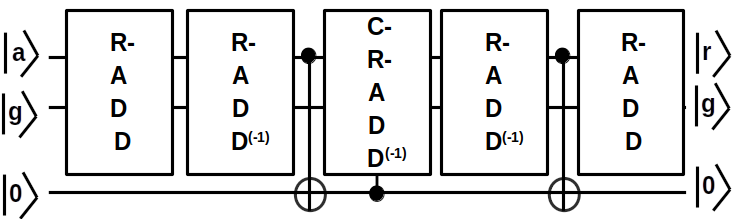}
    \caption{Registers $a$ (quantum) and $c$ (classical) are the two operands. $N$ is the ring size. $\ket{a}$ is the quantum register holding $a$ at the beginning of the circuit. The size of $a$ is $n+1$ because it will hold the sum $a+c$. $\ket{g}$ is a single garbage qubit used in the recursive adder, and $\ket{0}$ is marking the qubit controlling the addition/subtraction operations.}
    \label{fig:hma}
\end{figure}

The original circuit from \cite{vedral1996quantum} (Figure~\ref{fig:barenco}) has width $4n+O(1)$ where $3n+1$ of which are used to store $a, b$ and $N$ (which were not hardwired). Implementing constant-modular addition requires a single $n$-qubit quantum register, namely $\ket{a}$, while $c$ (we use $c$ instead of $b$ to highlight that it is a constant) and the size of the ring, $N$, are classical values. The modular addition is performed in the following steps (Figure~\ref{fig:hma}):
\begin{enumerate}
    \item Add $a$ and $c$.
    $\ket{a}\ket{g}\ket{0} \xrightarrow{add(a,c)} \ket{a+c}\ket{g}\ket{0}$
    
    \item Subtract $N$ from the previous sum, by running the recursive adder in inverse with $a+c$ and $N$ as integer and classical constant inputs respectively.
    
    \item If the flag bit of the result of the previous subtraction equals $1$ then it is negative and we need to re-add $N$. Otherwise, if it is positive, we leave it as is. 
    
    \item Reset the flag qubit to its original state. Subtract $c$ from $a+c\;mod(N)$. If it is $1$ (positive) then the flag qubit should be flipped. 
    
    \item Add $c$ to the result to recover $a+c\; mod(N)$.
\end{enumerate}

After the second step the state is $\ket{a+c-N}\ket{g}\ket{0}$. The state of the most significant qubit (MSB) of $\ket{a+c-N}$ indicates whether it is positive or negative. 

During the third step, if $a+c-N$ is indeed positive, then its MSB is $\ket{0}$ and $a+c-N=a+c\; mod(N)$. On the other hand, if $a+c-N<0$, the MSB is $\ket{1}$ and we should re-add the constant $N$. To implement both conditions in the circuit, we apply a $CNOT$ gate between MSB and the flag qubit which is initialised to $\ket{0}$. As a result, the flag equals $\ket{0}$ if the $a+c \geq 0$ and equals $\ket{1}$ in the other case. During the third step, we hence apply the recursive adder \emph{controlled} by the flag qubit adder with $a+c-N$ and $N$ as operands.

The fourth step resets the lowest qubit in Figure~\ref{fig:hma}, the flag qubit, to its original state $\ket{0}$. We subtract $c$ from $a+c \;mod(N)$ by applying the inverse of the recursive adder. If the result is positive, the flag qubit will be flipped: apply a CNOT between the most significant qubit of the result $a+c - c \; mod(N)$.

Effectively, the constant modular adder using the method from \cite{vedral1996quantum} has the width of the adder used as a component. Using the recursive addition circuit, the total width of the resulting modular adder is $n+3$. The input $\ket{a}$ is $n+1$ qubits wide in order to store carry. There are also two ancillae: 1) an ancilla to control the incrementation procedures all along the recursive addition operation, 2) the flag qubit from within the modular adder.

\section{Results and Resource Analysis}
\label{sec:analysis}

The presented modular adder was implemented in QUANTIFY~\cite{oumarou2020quantify} which is open-sourced and available at \url{https://github.com/quantumresource/quantify}.  We exhaustively tested the compiled Toffoli formulation of the adder using the Toffoli circuit simulator from \cite{toffolisim}.

Herein we focus on the Clifford+T cost of the adder, because the Toffoli gate decomposition influences the resource efficiency of the compiled circuit. The resource analysis was implemented with QUANTIFY, too.

To determine the depth of our modular adder, we need first to determine the depth of the Incrementer and the recursive adder. For the Incrementer, let $D_{Tf}$ denote the depth of a decomposed Toffoli, and $A$ present the number of ancillae, then for $n>1$ we have the following \emph{[D]epth} and \emph{[W]idth}:
\begin{align*}
    D_{Inc}(n) &= (6n-4)D_{Tf} + 2n -4 \\
    W(n) &= 2n+1+A
\end{align*}

The recursive adder is built with two components, namely the Incrementer and the Carry gate. Unlike the Incrementer, the depth of the Carry gate, and consequently the recursive adder, depends on the value of the constant $c$. In the following, we will study the worst case scenario that yields a maximum depth and which corresponds to $c$ being equal to $2^n-1$. Using the same notation as from the previous equation, the depth of the Carry gate for $n>2$ is given by:

\begin{align*}
D^{max}_{carry}(n) &=
    \begin{cases}
      (4n-6)D_{Tf}+4n-2, & \text{if}\; n\geq 3 \\
      4, & \text{if}\; n=2\\
      1, & \text{if}\; n=1
    \end{cases}
\end{align*}

Hence the depth of the recursive adder equals:
\begin{align*}
    D_{RA} &= 2\sum_{i=1}^{log(n)}(D_{Inc}(\frac{n}{2^i})+D^{max}_{carry}(\frac{n}{2^i})+ 2)\\
    W &= n + 2
\end{align*}

Lastly, our modular adder being composed of four recursive adders, one controlled version (which has the same depth but two more Toffoli gates that replace two CNOT gates) and two CNOT gates has the following depth:
\begin{align}
    D_{HBA} = 4\times D_{RA} + D_{CRA} + 2
\end{align}
with $D_{CRA}$ being the depth of the controlled version and is equal to $D_{CRA}=(4n-4)D_{Tf}+4n-4$.

The adder is of the ripple-carry type, and all the Toffoli gates are sequential. When using a Toffoli decomposition (see Appendix) which requires ancillae, only a constant number of ancillae would be needed since they can be reused for the rest of the Toffoli gates. The same fact applies for the Carry gate.

\section{Conclusion}

We designed a constant modular adder that uses only n+3 qubits which is a $~50\%$ reduction compared to the state-of-the-art constant modular adder from \cite{haner2017factoring} while simultaneously maintaining the linearity in the depth. We conjecture that the addition method from \cite{vedral1996quantum} is a generalisation of the incrementation trick (Section~\ref{sec:inc}).

We implemented the new constant modular adder in QUANTIFY\cite{oumarou2020quantify}, which is an open-sourced framework. Moreover, using this framework, we compiled the circuits using Toffoli gates and exhaustively verified the correctness. Future work is to use our modular addition for improving circuits from \cite{rines2018high}.

The proposed adder is useful for implementing, for example, quantum random walks or any other computation where state changes are a function of a constant. Our adder will also be useful for verifying very large quantum circuits that include constant modular addition. The size of the state-vectors and the overall matrix representing the circuit is reduced to half and quarter approximately when compared to \cite{haner2017factoring} and \cite{vedral1996quantum} respectively. Such reduction will have a quadratic and quatric speedup on the simulation of the quantum circuits. 

\bibliographystyle{plain}
\bibliography{main}

\begin{thebibliography}{10}

\bibitem{beauregard2003circuit}
Stephane Beauregard.
\newblock {Circuit for Shor's algorithm using 2n+ 3 qubits}.
\newblock {\em Quantum Information \& Computation}, 3(2):175--185, 2003.

\bibitem{cuccaro2004new}
Steven~A Cuccaro, Thomas~G Draper, Samuel~A Kutin, and David~Petrie Moulton.
\newblock A new quantum ripple-carry addition circuit.
\newblock {\em arXiv preprint quant-ph/0410184}, 2004.

\bibitem{draper2000addition}
Thomas~G Draper.
\newblock Addition on a quantum computer.
\newblock {\em arXiv preprint quant-ph/0008033}, 2000.

\bibitem{draper2006logarithmic}
Thomas~G Draper, Samuel~A Kutin, Eric~M Rains, and Krysta~M Svore.
\newblock A logarithmic-depth quantum carry-lookahead adder.
\newblock {\em Quantum Information \& Computation}, 6(4):351--369, 2006.

\bibitem{toffolisim}
Casey Duckering.
\newblock {Cirq Toffoli circuit simulator}.
\newblock
  \url{https://github.com/cduck/cirqtools/blob/master/cirqtools/classical_simulator.py}.

\bibitem{fowler2004implementation}
AG~Fowler, SJ~Devitt, and LCL Hollenberg.
\newblock Implementation of shor's algorithm on a linear nearest neighbour
  qubit array.
\newblock {\em Quantum Inf. Comput.}, 4(quant-ph/0402196):237--251, 2004.

\bibitem{haner2017factoring}
Thomas H{\"a}ner, Martin Roetteler, and Krysta~M Svore.
\newblock {Factoring using 2n+ 2 qubits with Toffoli based modular
  multiplication}.
\newblock {\em Quantum Information \& Computation}, 17(7-8):673--684, 2017.

\bibitem{munoz2018quantum}
Edgard Mu{\~n}oz-Coreas and Himanshu Thapliyal.
\newblock {Quantum circuit design of a T-count optimized integer multiplier}.
\newblock {\em IEEE Transactions on Computers}, 68(5):729--739, 2018.

\bibitem{oonishi2020efficient}
Kento Oonishi, Tomoki Tanaka, Shumpei Uno, Takahiko Satoh, Rodney Van~Meter,
  and Noboru Kunihiro.
\newblock {Efficient Construction of a Control Modular Adder on a
  Carry-Lookahead Adder Using Relative-phase Toffoli Gates}.
\newblock {\em arXiv preprint arXiv:2010.00255}, 2020.

\bibitem{oumarou2020quantify}
Oumarou Oumarou, Alexandru Paler, and Robert Basmadjian.
\newblock {QUANTIFY: A framework for resource analysis and design verification
  of quantum circuits}.
\newblock In {\em 2020 IEEE Computer Society Annual Symposium on VLSI
  (ISVLSI)}, pages 126--131. IEEE, 2020.

\bibitem{rines2018high}
Rich Rines and Isaac Chuang.
\newblock High performance quantum modular multipliers.
\newblock {\em arXiv preprint arXiv:1801.01081}, 2018.

\bibitem{njross}
Neil~J. Ross and Peter Selinger.
\newblock Optimal ancilla-free clifford+t approximation of z-rotations.
\newblock {\em Quantum Info. Comput.}, 16(11–12):901–953, September 2016.

\bibitem{vedral1996quantum}
Vlatko Vedral, Adriano Barenco, and Artur Ekert.
\newblock Quantum networks for elementary arithmetic operations.
\newblock {\em Physical Review A}, 54(1):147, 1996.

\end{thebibliography}

\section*{Appendix}

\begin{figure}
    \centering
    \includegraphics[width=\columnwidth]{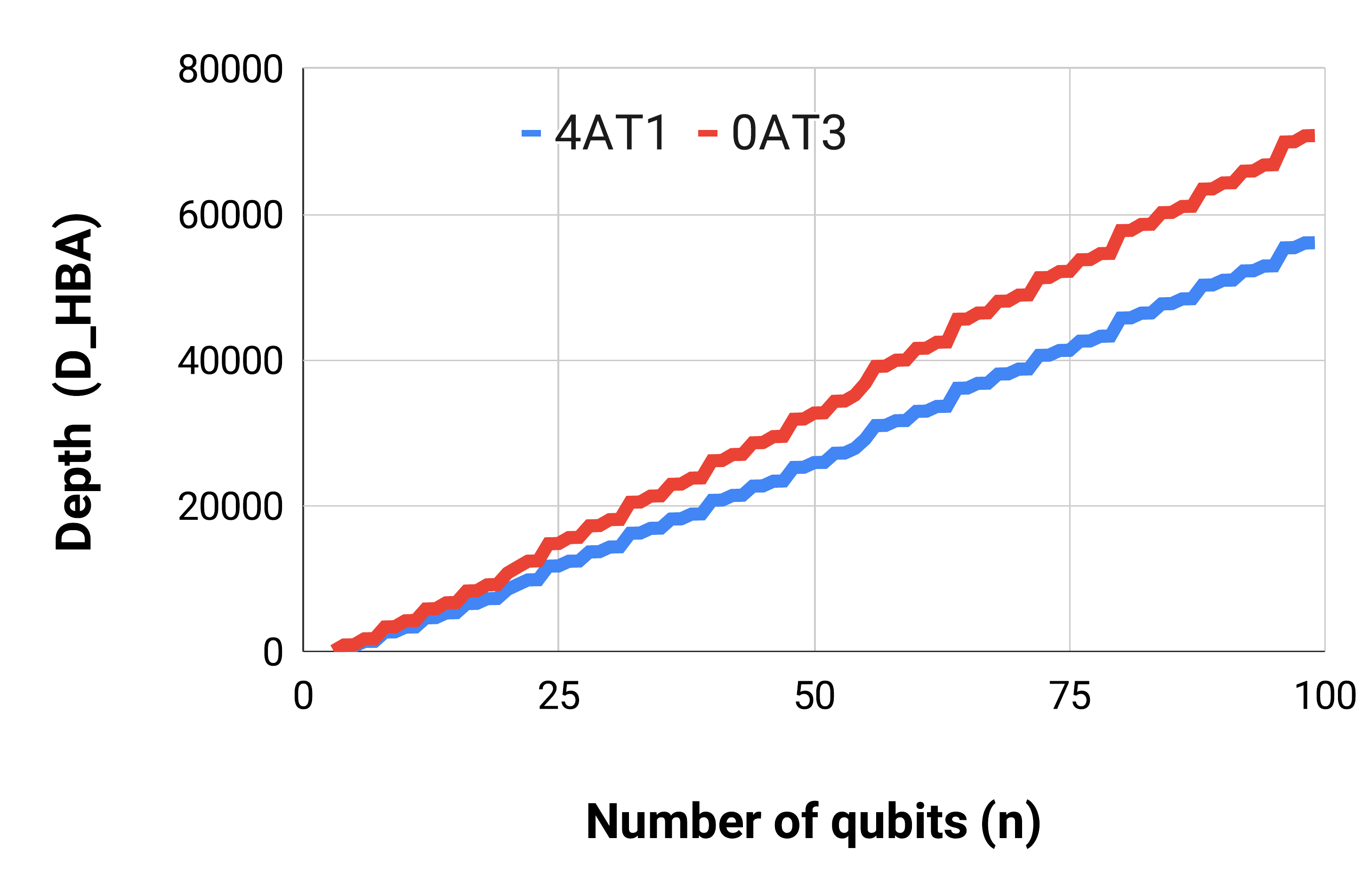}
    \caption{The 4AT1 Toffoli decomposition offers better depth than the 0AT3 decomposition.}
    \label{fig:kq}
\end{figure}

\begin{figure}[!h]
    \centering
    \includegraphics[width=\columnwidth]{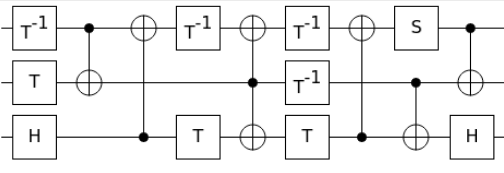}
    \caption{The 0AT3 Toffoli gate decomposition uses no ancillae and has T-depth 3.}
    \label{fig:toff_0at3}
\end{figure}

\begin{figure}[!h]
    \centering
    \includegraphics[width=\columnwidth]{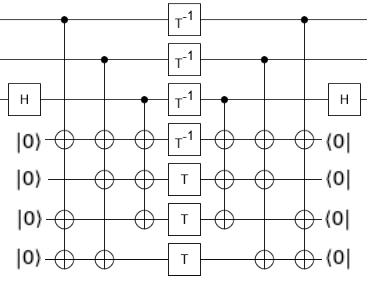}
    \caption{The 4AT1 Toffoli gate decomposition uses four ancillae and has T-depth 1.}
    \label{fig:toff_4at1}
\end{figure}

Figure~\ref{fig:kq} illustrates the depth of the modular adder when using the 4AT1 (Figure~\ref{fig:toff_4at1}) and the 0AT3 (Figure~\ref{fig:toff_0at3}) decompositions. The four ancillae from 4AT1 are reused for all the Toffoli gates of the Incrementer, such that the overall area is better when using 4AT1.

Concerning the recursive adder within the modular adder, three scenarios exist regarding the decomposition of the Toffoli gates. We consider the $\mathcal{O}(nlog(n))$-depth sequential recursive adder and the $\mathcal{O}(n)$-depth parallel recursive adder:

\begin{enumerate}
    \item Using the sequential adder and without any parallelism: all Toffoli gates are decomposed using 4AT1 such that only 4 ancillae are introduced and can be reused for the recursive addition sub-circuits.
    \item Using the parallel adder and maintaining the parallelism: we choose the 4AT1 decompositions and the parallel recursive addition sub-circuits use different ancillae, for a total of $4\times \frac{n}{4}=n$ -- because the leaves do not have Toffoli gates \footnote{There are n/2 additions of single bits, n/4 additions of 2 bits, etc. Single bit additions use CNOTs. Toffoli gates are used only starting with the 2 bit additions.}.
    
    \item Using the parallel adder without introducing ancillae: the Toffoli gates are decomposed using 0AT3. The depth increases by a factor of $9/7$ if parallel CNOTs are allowed, otherwise the depth is actually reduced by $19/9 \approx 2$.
\end{enumerate}

The gate count of the recursive adder circuit is in $\mathcal{O}(nlog(n))$.  If gate parallelism is allowed, the depth can be reduced to $\mathcal{O}(n)$, with two options (cf. Fig.~\ref{fig:recadd}):
\begin{enumerate}
    \item With $\frac{n}{2}$ ancilla. Introduce as many as necessary ancillae to use them as garbage qubits and hence parallelise the sub-circuit blocks in each recursion. This would mean that $\frac{n}{2}$ ancillae are added in total.
    
    \item Without ancilla. Without loss of generality, we only execute the sub-circuit blocks on the lower half $\ket{a_L}$. We then use the qubits of the upper half $\ket{a_H}$ as a garbage qubits to parallelise the sub-circuit blocks in each recursion round. Once the recursion is finished on the lower half $\ket{a_L}$, we execute the sub-circuit blocks on the upper half $\ket{a_H}$ using $\ket{a_L}$ as garbage qubits for the same purpose.
\end{enumerate}

Without parallelism, the area (depth x width) of the adder is definitely worse than with parallelism. We have an $\mathcal{O}(nlog(n))$ depth and still a linear width, even though no ancillae were introduced with 0AT3. As a result, the overall area is equivalent to $\mathcal{O}(n^2log(n))$.

When parallelising the adder and using 4AT1, the area scales asymptotically in $\mathcal{O}(n^2)$. However, with 4AT1 there is a depth ratio of $\sim \frac{7}{9}$ compared to the third option with 0AT3 but with approximately twice as much width. As a result, when determining the area, the third decomposition scenario with the 0AT3 decomposition is better than the second alternative with the 4AT1 decomposition. Because, the depth ratio $\sim \frac{7}{9}$ is less than the width ratio $\sim 2$. 

The overall area of the adder decomposed with 4AT1 in the last scenario is then $\sim \frac{7}{9}\times 2=\frac{14}{9} > 1$ bigger than that when 0AT3 is used.

\end{document}